%
%
\def\Cov{{\rm Cov}}

\catcode`\@=11
\def\xtitleb#1#2{\par\stepc{Tm}
    \resetcount{Tn}
    \if N\lasttitle\else\vskip\tbbeforeback\fi
    \bgroup
       \normalsize
       \raggedright
       \pretolerance=10000
       \it
       \setbox0=\vbox{\vskip\tbbefore
          \normalsize
          \raggedright
          \pretolerance=10000
          \noindent \it #1.\arabic{Tm}.\ \ignorespaces#2
          \vskip\tbafter}
       \dimen0=\ht0\advance\dimen0 by\dp0\advance\dimen0 by 2\baselineskip
       \advance\dimen0 by\pagetotal
       \ifdim\dimen0>\pagegoal
          \ifdim\pagetotal>\pagegoal
          \else \if N\lasttitle\eject\fi \fi\fi
       \vskip\tbbefore
       \if N\lasttitle \penalty\subsection@penalty \fi
       \global\subsection@penalty=-100
       \global\subsubsection@penalty=10007
       \noindent #1.\arabic{Tm}.\ \ignorespaces#2\par
    \egroup
    \nobreak
    \vskip\tbafter
    \let\lasttitle=B%
    \parindent=0pt
    \everypar={\parindent=\stdparindent
       \penalty\z@\let\lasttitle=N\everypar={}}%
       \ignorespaces}
\def\xautnum#1{\global\advance\eqnum by 1\relax {\rm (#1\the\eqnum)}}
\catcode`\@=13
%
%
%
  \MAINTITLE={ Improving the accuracy of mass reconstructions from
  weak lensing: local shear measurements}
  \SUBTITLE={ ????? } 
  \AUTHOR={ Marco Lombardi and Giuseppe Bertin } 
  \OFFPRINTS={ M. Lombardi } 
  \INSTITUTE={ Scuola Normale Superiore, Piazza dei Cavalieri 7, I
  56126 Pisa, Italy }
  \DATE={ ????? } 
  \ABSTRACT={ Different options can be used in order to measure the
  shear from observations in the context of weak lensing. Here we
  introduce new methods where the isotropy assumption for the
  distribution of the source galaxies is implemented directly on the
  observed quadrupole moments. A quantitative analysis of the error
  associated with the finite number of source galaxies and with their
  ellipticity distribution is provided, applicable even when the shear
  is not weak. Monte Carlo simulations based on a realistic sample of
  source galaxies show that our procedure generally leads to errors
  $\approx 25 \%$ smaller than those associated with the standard
  method of Kaiser and Squires (1993). } 
  \KEYWORDS={ gravitational lensing -- dark matter -- galaxies:
  clustering } 
  \THESAURUS={ 12.07.1; 12.04.1; 11.03.1 }
  \maketitle
  \MAINTITLERUNNINGHEAD{ Accurate shear from weak lensing }
  \AUTHORRUNNINGHEAD{ M. Lombardi \& G. Bertin }
%
%

\titlea{Introduction}
One of the most interesting applications of gravitational lenses
arises in the context of weak distortions of far sources (e.g., see
Tyson {\it et al}.\ 1984, Webster 1985). The presence of a
gravitational lens (e.g., an intervening cluster of galaxies) breaks
the symmetry of the image of an isotropic population of extended
sources, with the lensed objects preferentially oriented in the
tangential direction with respect to the center of the lens. This
effect is usually quantified in terms of the {\it shear} introduced by
the lens. In turn, the shear field measured from the observed
anisotropy and stretching in a field of distant objects can be used to
infer the two-dimensional (projected) mass distribution of the lens
(see Kaiser and Squires 1993; Kaiser {\it et al}.\ 1995).

At present these concepts have found application in two different
classes of lenses, i.e. galaxies and clusters. The work on galaxies
has been performed via a statistical investigation of a suitable
ensemble (Griffiths {\it et al}.\ 1996, Brainerd {\it et al}.\ 1996),
due to the limited number of background sources generally associated
with an individual galaxy. Relatively nearby clusters, instead, are
ideal candidates for direct weak lens analyses (e.g., Kneib {\it et
al}.\ 1996; for a review, see Kneib and Soucail 1995; see also Gould
and Villumsen 1994); these studies are especially important in
relation to the problem of dark matter on the megaparsec scale, with
significant cosmological implications.

In this paper we focus on the measurement of shear from observations,
as a key step in the process of estimating the mass of a cluster of
galaxies from weak gravitational lensing. We discuss the possible
merits of alternative options in handling a given set of data under
the same conditions for the population of background
sources. Different methods may introduce different forms of bias,
depending on whether the data are degraded by seeing or by other
effects. In this article we give special attention to one aspect of
the problem which is largely independent of the specific
characteristics of the observations involved. This important factor in
determining the size of the {\it error} in the shear measurement is
the ellipticity distribution of the source galaxies. In Sect.~2, with
a short discussion of the ``standard" method of Kaiser and Squires
(1993), we provide an expression for the expected error in a shear
measurement as a function of the ellipticity distribution of the
sources. This result is not restricted to a weak shear analysis. Then,
in Sect.~3, we introduce a different method, directly based on the
observed quadrupole moments, which retains information not only on the
shape of the observed objects, but also on their angular size, and we
carry out the related error analysis. The method is shown to implement
correctly the hypothesis of a random orientation of the sources; it is
also shown to be fully equivalent to the ``standard'' method if
applied to a population of nearly round source objects. The method is
then generalized and further improved by introducing suitable weight
functions to optimize the inversion procedure. In Sect.~4 we describe
the results of a wide set of Monte Carlo simulations applied to
realistic distributions of ellipticities for the source
galaxies. These show that the new method proposed in this paper can be
significantly more accurate than the standard method in determining
the shear from the data. The simulations also confirm that the
applicability of the method and the conclusions on its accuracy do not
depend on the strength of the lens in a relatively wide parameter
range.

\titlea{The isotropy hypothesis and the ``standard'' method}
\titleb{Notation}

Following standard notation with minor modifications (see Schneider et
al.\ 1992), let $\vec{\theta}^{\rm s}$ be the unlensed position of a
point source and $\vec{\theta}$ the position of the corresponding
observed image. We will generally refer to a Cartesian representation
of the {\it angle\/} vectors, so that $\vec{\theta} = (\theta_1,
\theta_2)$. The deflection angle $\vec{\beta}$ is defined through the
relation
$$
\vec{\theta}^{\rm s} = \vec{\theta} - \vec{\beta} (\vec{\theta}) \; .
\eqno\autnum
$$
For a single thin lens we can define $D_{\rm os}$, $D_{\rm ds}$, and
$D_{\rm od}$ as the distances between the observer (o), the lens or
deflector (d), and the source (s). In this case the deflection angle
can be expressed as a function of the dimensionless mass distribution
$\kappa(\vec{\theta}) = \Sigma(\vec{\theta}) / \Sigma_{\rm c}$, where
$\Sigma(\vec{\theta})$ is the projected mass distribution of the
deflector and $\Sigma_{\rm c} = c^2 D_{\rm os} / ( 4 \pi G D_{\rm ds}
D_{\rm od})$ is the critical density; it is given by
$$
\vec{\beta}(\vec{\theta}) = {1 \over \pi} \int {\kappa(\vec{\theta}')
(\vec{\theta} - \vec{\theta}') \over \| \vec{\theta} - \vec{\theta}'
\|^2 } \,\diff^2 \theta' \; .
\eqno\autnum
$$
Thus for a small extended source in the vicinity of $\vec{\theta} =
\vec{\theta}_0$ we have $\vec{\theta}^{\rm s} \simeq \vec{\theta}^{\rm
s} (\vec{\theta}_0) + J(\vec{\theta}_0) ( \vec{\theta} -
\vec{\theta}_0 )$, with
$$\eqalignno{
J(\vec{\theta}) &= \left( {\partial \vec{\theta}^{\rm s} \over
\partial \vec{\theta}} \right) & \cr
&= \left( \matrix{ 1 - \kappa(\vec{\theta}) + \gamma_1(\vec{\theta}) &
\gamma_2(\vec{\theta}) \cr \gamma_2(\vec{\theta}) & 1 -
\kappa(\vec{\theta}) - \gamma_1(\vec{\theta}) \cr } \right) & \cr
&= \bigl(1 - \kappa(\vec{\theta}) \bigr) \left( \matrix{ 1 +
g_1(\vec{\theta}) & g_2(\vec{\theta}) \cr g_2(\vec{\theta}) & 1 -
g_1(\vec{\theta}) } \right) \; .& \autnum \cr}
$$
The complex quantity $\gamma = \gamma_1 + \imag \gamma_2$ is called
shear because it is responsible for the distortion of the image. The
reduced shear parameter is defined as $g = g_1 + \imag g_2 = \gamma /
(1 - \kappa)$ and is the quantity actually derived from the
observations; note that for a weak lens $g \simeq \gamma$.

From the surface brightness distribution $I(\vec{\theta})$ of an
extended object we can define the total luminosity $I$, the position
vector of the center of the image $\vec{\delta}$, and the image quadrupole
$Q_{ij}$ as
$$\eqalignno{
I &= \int I(\vec{\theta}) \,\diff^2 \theta \; , & \autnum \cr
\vec{\delta} &= {1 \over I} \int \vec{\theta} I(\vec{\theta}) \,\diff^2
\theta \; ,& \autnum \cr
Q_{ij} &= {1 \over I} \int (\theta_i - \delta_i)(\theta_j - \delta_j)
I(\vec{\theta}) \,\diff^2 \theta \; . & \autnum \cr}
$$
The quadrupole $Q_{ij}$ gives information on angular size, shape, and
orientation of the observed galaxy. If we are not interested in the
image size, we may refer to the (complex) ellipticity $\chi$:
$$
\chi = \chi_1 + \imag \chi_2 = {Q_{11} - Q_{22} + 2 \imag Q_{12} \over
Q_{11} + Q_{22}} = \chi(Q) \; .
\eqno\autnum
$$
The modulus of $\chi$ is related to the shape ($|\chi| = 0$ for a
circular object) while its argument is twice the image orientation
angle (angle between the $\theta_1$ axis and the galaxy major axis).

Similar definitions can be given for the unlensed source, with
$I(\vec{\theta})$ replaced by the source luminosity flux $I^{\rm
s}\bigl( \vec{\theta}^{\rm s} \bigr)$. For a small object a simple
relation holds between the observed and the source quadrupole moment:
$$
Q^{\rm s} = J Q J \; ,
\eqno \autnum
$$
where we have used the symmetry of $J$. The corresponding relation
between $\chi$ and $\chi^{\rm s}$ is (here an asterisk denotes complex
conjugation)
$$
\chi^{\rm s} = {\chi + 2 g + \chi^* g^2 \over 1 + 2 \Re \bigl(
\chi^* g \bigr) + |g|^2} = \chi^{\rm s}(\chi, g) \; .
\eqno \autnum
$$
Thus $\chi^{\rm s}$ is a function of $\chi$ and $g$ only.

\titleb{The standard method (X method)}

The local value of $g$ can be derived from the observation of a large
number $N$ of galaxies in a small patch of the sky. For the purpose,
it is usually assumed that the population of source objects is {\it
isotropic}, i.e. that the source galaxies have random
orientations. The area of the sky under consideration must be small so
that the Jacobian matrix can be considered to be approximately
constant, i.e. the same for all the galaxies observed in the area. The
source galaxies are taken to have $D_{\rm os}/D_{\rm ds} \simeq
\hbox{constant}$, which is a reasonable assumption for far away sources
lensed by a nearby cluster.

The standard method (X method) of Kaiser and Squires (1993) implements
the isotropy hypothesis by arguing that the quantity
$$
\widetilde{\chi^{\rm s}} = {1 \over N} \sum_{n = 1}^N \chi^{\rm
s}  \bigl(\chi^{(n)}, g\bigr)
\eqno \autnum
$$
vanishes approximately. For a given set of data $\big\{ \, \chi^{(n)}
\, \big\}$, with $n = 1, \dots , N$, the condition
$$
\widetilde{\chi^{\rm s}} = 0
\eqno \autnum
$$
is used to determine the reduced shear parameter $g$. In terms of the
probability distribution for the {\it source\/} ellipticities
$p\big(\chi^{\rm s} \big)$, the isotropy hypothesis states that
$p\big(\chi^{\rm s} \big) = p\big( \big| \chi^{\rm s} \big|
\big)$. Thus we have $\bigl\langle \chi^{\rm s} \bigr\rangle = 0$,
with a scalar covariance matrix $\bigl\langle \chi^{\rm s}_i \chi^{\rm
s}_j \bigr\rangle = c \delta_{ij}$. [Here the indices refer to the
representation of the complex shear as $\chi = \chi_1 + \imag
\chi_2$.] Given the range of values for $| \chi |$, we must have $c
\le 1/2$. Then the central limit theorem ensures that the variable
$\widetilde{\chi^{\rm s}}$ of Eq.~(10), calculated for $N$ galaxies,
must converge to a Gaussian with mean $0$ and covariance matrix $(c /
N) \delta_{ij}$ in the limit $N \gg 1$.

\titleb{Error analysis for the standard method}
Here we consider the simple case of a {\it sharp\/} distribution of
source ellipticities, when the population of objects is basically made
of nearly round source galaxies. We then assume that the probability
distribution $p$ is characterized by $c \ll 1$. For such
distributions, the condition $\widetilde{\chi^{\rm s}} = 0$ is shown
to give a correct determination of the reduced shear parameter $g$
(see Appendix~A). If we perform a measurement for a lens with true
reduced shear equal to $g_0$, then the expected statistics for $g$ has
mean and covariance given by
$$\eqalignno{
& \langle g \rangle = g_0 \; ,& \autnum \cr
& \Cov_{ij} (g) = {c \bigl( 1 - |g_0|^2 \bigr)^2 \over 4 N} \delta_{ij}
\; . & \autnum \cr}
$$
Here and in the following, by $\Cov_{ij}(X) = \bigl\langle (X_i -
\overline{X}_i) (X_j - \overline{X}_j) \bigr\rangle $ we denote the
covariance matrix for the random variable $X$ with mean $\overline{X}
= \langle X \rangle$.  The expected error, $\sigma_g = \bigl| 1 -
|g_0|^2 \bigr| \sqrt{c / 4 N}$, is the same on $g_1$ and on $g_2$. The
result stated in Eq.~(12) has been obtained also by Schneider \& Seitz
1995 (see also Schramm \& Kayser 1995); expression~(13) is consistent
with the results by Miralda-Escud\'e (1991) and by Schneider \& Seitz
(1995), who, however, do not bring out the $\bigl( 1 - |g_0|^2 \bigr)$
dependence.

The error depends on the source distribution via the parameter $c$ and
shows that the method can lead to an accurate measurement of $g$
provided $N$ is sufficiently large. The error analysis in the general
case of a broad distribution ($c \la 1/2$) is much more
difficult. However, if we naively extrapolate the conclusions obtained
for sharp distributions, the largest error on $g_1$ and on $g_2$
should be $\bigl| 1 - |g_0|^2 \bigr| / \sqrt{8 N}$. In the opposite
limit, when $c = 0$, the error vanishes. In fact, when $c=0$ all
galaxies are circular. In this situation the value of $g_0$ can be
derived simply by observing the shape of a single galaxy, and thus the
error on $g$ is zero.

The dependence of the error on $g_0$ is quite interesting. For $|g_0|
< 1$ the error decreases with increasing $|g_0|$, and it vanishes when
$|g_0| = 1$. This happens because for $|g_0| = 1$ the lens is
singular, since $\det J \propto \bigl( 1 - |g_0|^2 \bigr) = 0$, which
means that all galaxies are seen as thin segments, with $| \chi | =
1$. If the source population is made of nearly round galaxies, by
observing even a single galaxy with very high ellipticity, we can
immediately infer that $|g_0| \simeq 1$; of course the argument of
$g_0$ is given by the orientation of the galaxy. The behavior of the
error for $|g_0| > 1$ is explained in terms of the $g_0 \mapsto
1/g_0^*$ local invariance (Schneider \& Seitz 1995, Seitz \& Schneider
1995). Thus it should be possible to infer the error for a value of
$g_0$ with modulus $|g_0| > 1$ from the error for $g_0' = 1/g_0^*$. In
fact (see Appendix~A), the {\it relative\/} errors for $g_0$ and $g_0'$
are the same.

\titlea{Alternative methods based on the observed average quadrupole}

\titleb{Simple quadrupoles (Q method)}
We now introduce a new method based on the observed quadrupole moments
rather than on ellipticities (Q method). In order to implement the
isotropy assumption we replace condition (11) by
$$
\chi^{\rm s} \biggl( {1 \over N} \sum_{n=1}^N Q^{(n)} , g \biggr) = 0
\; .
\eqno \autnum
$$
Note that here the average is performed {\it inside the parentheses},
i.e.\ on the observed quadrupoles. Then it is argued that the average
quadrupole is traced back to a {\it circular source}. The method is
thus similar to the one used by Bonnet \& Mellier (1995), but with an
important difference in that we refer to an average on quadrupole moments,
which gives different weights to galaxies with different angular
sizes (see additional comments at the end of Sect.~3.2).

To see why Eq.~(14) holds, we define the probability distribution for
the {\it source\/} quadrupole moments $p_Q \bigl( Q^{\rm s}
\bigr)$. Since the quadrupole matrix $Q^{\rm s}_{ij}$ is symmetric,
$p_Q \bigl( Q^{\rm s} \bigr)$ is a function of three real quantities
${\cal Q}^{\rm s}_1 = Q^{\rm s}_{11}$, ${\cal Q}^{\rm s}_2 = Q^{\rm
s}_{12} = Q^{\rm s}_{21}$, and ${\cal Q}^{\rm s}_3 = Q^{\rm
s}_{22}$. The isotropy hypothesis states that the probability
distribution $p_Q$ is independent of the orientation of the galaxy, so
that $\bigl\langle Q^{\rm s} \bigr\rangle = M {\rm\ Id}$, with $\rm
Id$ the $2 \times 2$ identity matrix. Furthermore, isotropy requires
the covariance matrix of ${\cal Q}^{\rm s}$ to be of the form
$$
\Cov_{kl} \bigl( {\cal Q}^{\rm s} \bigr) = \left( \matrix{ d + e & 0 &
d - e
\cr 0 & e & 0 \cr d - e & 0 & d + e \cr} \right) \; .
\eqno \autnum
$$
As the covariance matrix is positive definite, we must have $d, e \ge
0$. Here the two indices $k$ and $l$ run on the three independent
components of ${\cal Q}^{\rm s}$. It is possible to show that $e$ is
related to the constant $c$ defined earlier in Sect.~2.2, as $c = e /
M^2$, while $d$ is the variance of $Q^{\rm s}_{11} + Q^{\rm s}_{22}$,
the size of the galaxy.

If we have a large number $N$ of galaxies with quadrupole moments
$\bigl\{ \, Q^{(n)} \, \bigr\}$, with $n = 1, \dots , N$, we can write
$$
J \biggl( {1 \over N} \sum_{n = 1}^N Q^{(n)} \biggr) J = {1 \over N}
\sum_{n = 1}^N Q^{{\rm s}(n)} \simeq M {\rm\ Id}\; ,
\eqno \autnum
$$
where the approximate sign is used because $N$ is finite. Let us now
calculate the ellipticity $\chi$ for both sides of this equation by
applying the definition given by Eq.~(7)
$$
\chi \Biggl( J \biggl( {1 \over N} \sum_{n = 1}^N Q^{(n)} \biggr) J
\Biggr) \simeq \chi( M {\rm\ Id} ) = 0\; .
\eqno \autnum
$$
Equation (17) is the result stated in Eq.~(14), because $\chi(JQJ) =
\chi^{\rm s}(Q, g)$.

The reduced shear $g$ calculated through Eq.~(14) is a good estimate
of the true reduced shear $g_0$. In Appendix~A we show that in the
case of sharp distributions the quantity $g$ has mean and covariance
matrix given by
$$\eqalignno{
\langle g \rangle &= g_0 \; , & \autnum \cr
\Cov_{ij} (g) &= {c \bigl( 1 - |g_0|^2 \bigr)^2 \over 4 N} \delta_{ij}
\; . & \autnum \cr}
$$
Note that these expressions are exactly the same as for the standard
method.  This apparently surprising result can be traced to the
decoupling of size and ellipticity distributions for quasi-circular
sources. This property is shown directly by Eq.~(15), where
$\Cov\bigl( {\cal Q}^{\rm s} \bigr)$ is found to depend on two
quantities, $d$ being related to the size distribution and $e$ being
related to the ellipticity.

\titleb{Weighted quadrupoles (W method)}
The simple quadrupole method described in Sect.~3.1 is based on the
isotropy requirement $\bigl\langle Q^{\rm s} \bigr\rangle = M {\rm\
Id}$, which leads to Eq.~(16) and thus justifies the use of
Eq.~(14). Since the analysis has shown that $g$ is determined more
accurately when the ellipticities involved are smaller, we may argue
that a generalization where a {\it penalty\/} is assigned to galaxies
with large $\bigl| \chi^{\rm s} \bigr|$ should be able to improve the
method. Thus we introduce a weight function $W\bigl( Q^{\rm s} \bigr)
> 0$ with the following requirements:
\medskip
\item{1.} $W$ depends only on the {\it source\/} quadrupoles and is
invariant under rotation (consistent with the isotropy of $p_Q\bigl(
Q^{\rm s} \bigr)$).
\item{2.} If $k$ is a constant multiplying factor, $W\bigl( k
Q^{\rm s} \bigr) = f(k) W\bigl( Q^{\rm s} \bigr)$ (consistent with
the fact that the analysis of the data is able to constrain $J$ only
up to a multiplying factor $(1 - \kappa)$).
\item{3.} $W\bigl( Q^{\rm s} \bigr)$ penalizes objects with
large $\bigl| \chi^{\rm s} \bigr|$.
\medskip
Under these conditions we may now consider as the starting point the
relation $\bigl\langle Q^{\rm s} W\bigl( Q^{\rm s} \bigr) \bigr\rangle
= M_W {\rm\ Id}$, with $M_W$ a positive number dependent on $W$, which
holds because of the isotropy assumption. Then, following the argument
that has led to Eq.~(17), we find
$$\eqalignno{
\chi \Biggl( J \biggl( {1 \over N} \sum_{n=1}^N Q^{(n)} W \bigl( J
Q^{(n)} J \bigr) \biggr) J \Biggr) &\simeq \chi( M_W {\rm\ Id} ) \cr
&= 0 \; , & \autnum \cr}
$$
which indicates that we could replace Eq.~(14) by
$$
\chi^{\rm s} \biggl( {1 \over N} \sum_{n=1}^N Q^{(n)} W \bigl( J Q^{(n)} J
\bigr)
, g \biggr) = 0 \; .
\eqno \autnum
$$

In this method based on weighted quadrupoles (W method) one has to
proceed by iteration, because the quadrupoles $JQ^{(n)}J = Q^{{\rm
s}(n)}$ that appear in the weight functions {\it are not given
directly by the data\/} but must be guessed first in order to start
the iteration procedure. Several alternatives have been
considered. One possibility is to take the result of the method
described in Sect.~3.1 as the initial seed for the iteration. Another
reasonable choice is to take the natural approximation of the weak
lensing limit, i.e. $g \simeq - (1 / 2N) \sum_n \chi^{(n)}$, or the
estimate obtained from $g / \bigl( 1 + |g|^2 \bigr) = - (1 / 2N)
\sum_n \chi^{(n)}$ applicable beyond the weak lensing limit (for a
sharp ellipticity distribution; see Appendix~A).

As far as the choice of the weight functions is concerned, we have
found that a logarithmic dependence
$$
W\bigl( Q^{\rm s} \bigr) = - \ln \bigl( \bigl| \chi^{\rm s} \bigr|
\bigr)
\eqno \autnum
$$
is simple and suitable for the purpose.

Note that a weight function could be introduced also for the standard
X method. As indicated at the beginning of this subsection, the proper
way to impose the weights refers to the {\it source\/} quadrupole
moment $Q^{\rm s} = J Q J$ and not to the {\it observed\/} quantity
$Q$. Then Eq.~(10) would be replaced by
$$
\widetilde{\chi^{\rm s}} = {1 \over N} \sum_{n=1}^N W\bigl( J Q^{(n)} J
\bigr) \chi^{\rm s} \bigl(\chi^{(n)}, g \bigr) \; .
\eqno \autnum
$$
From this point of view, the Q method could thus be seen as a
special case of weighted X method, with weights
$$
W\bigl( Q^{\rm s} \bigr) = Q^{\rm s}_{11} + Q^{\rm s}_{22} \; ,
\eqno \autnum
$$
as can be checked from the relevant definitions. Similarly, the W
method can always be transformed into a weighted X method, and
vice-versa.

The weighted quadrupole method presents some analogies with the {\it
rejection\/} technique used by Bonnet \& Mellier (1995).  Galaxies
with large ellipticities are weakly distorted by the lens (with little
information on the shear), and contribute significantly to the
dispersion. Bonnet \& Mellier (1995) thus argue that only galaxies
with small ellipticities should be retained and galaxies with
ellipticities larger than a given threshold value $\chi_{\rm cut}$
should be rejected. Hence rejection can be considered as a particular
weighted method with a step-like weight function. However, the
rejection apparently used in the above-mentioned article refers to the
{\it observed} (rather than the {\it source\/}) ellipticities. This in
turn introduces an undesired anisotropy in the galaxy sample, and thus a
{\it bias\/} in the measured shear. This point is easily clarified in
the weak lensing approximation. In this case Eq.~(9) becomes simply
$\chi^{\rm s} = \chi + 2 \gamma_0$ so that the lens acts as a
translation on the ellipticity. A rejection on $|\chi|$ then becomes a
rejection on $| \chi^{\rm s} - 2 \gamma_0 |$, i.e.\ a rejection of all
galaxies that in the $\chi^{\rm s}$ complex plane are outside a circle
of radius $\chi_{\rm cut}$ centered on $2 \gamma_0$. Therefore, using
the X method, the expected value for $\gamma$ is reduced, i.e.
$$
\langle \gamma \rangle \simeq \gamma_0 - {\pi \over {\cal F}}
p(\chi_{\rm cut}) \chi_{\rm cut}^2 \gamma_0 \; ,
\eqno \autnum
$$
where $\cal F$ is the fraction of unrejected galaxies and $\chi_{\rm
cut}$ is the maximum allowed value for $| \chi |$. This expression is
valid for $\chi_{\rm cut} \le 1 - 2\gamma_0$ and to first order in
$\gamma_0$. Similar comments could be made on the rejection using the Q
method, but here $\langle \gamma \rangle$ depends in a more
complicated way on the probability distribution $p_Q$. In conclusion
a rejection referred to the observed ellipticities introduces a bias
on $\gamma$ in the direction of smaller values.

One might also try to relate the W~method to maximum-likelihood
methods used in this context (e.g., see Bartelmann {\it et
al}. 1996). The relative merits may depend on the characteristics of
the population of source galaxies (see also Sect.~4).

The arguments discussed after Eq.~(22) have found convincing
demonstration in the simulations outlined below in Sect.~4.

\titleb{Advantages with respect to the standard method}
The following is a short qualitative discussion of the three methods
introduced above in relation to the properties of the background
source galaxy population.

Our starting point is the fact that the expected error on $g$ is
proportional to $\sqrt{c}$, i.e.\ it is larger for broader
distributions. This suggests that the error is dominated by the
contribution of relatively flat source galaxies. This has led us to
introduce the weighted quadrupole method. On the other hand, the
simple quadrupole method should also perform better than the standard
method. [We should emphasize (see comment after Eq.~(19)) that the
apparent full equivalence of the X and Q methods suggested by
Eqs.~(12, 13) and (18, 19) holds only in the limit $c \rightarrow 0$,
where size and ellipticity decouple from each other, as shown by
Eq.~(15).]  This may be argued in the following way. Consider a simple
case where all source galaxies are flat disks with the same luminosity
and the same size. Then, for a random spatial orientation, all
galaxies will appear as ellipses, with identical major axes and minor
axes ranging from $0$ to the size of the major axes. Therefore flatter
source objects occupy smaller areas of the sky. In contrast with the
standard method, the quadrupole method takes the galaxy size into
account, which should lead to more accurate results. The argument can
be extended to the case where different luminosities and sizes are
involved. Furthermore, it should also be applicable to a more
realistic galaxy population, since galaxy fields are usually dominated
by spiral galaxies. This will be confirmed by our simulations (see
Sect.~4).

It is also interesting to consider the case of a source population
made of spheroids (i.e. axisymmetric ellipsoids). For a population of
oblate galaxies we expect a behavior similar to that of disk
galaxies. Thus if elliptical galaxies are generically oblate, their
contribution would not alter the argument in favor of the quadrupole
method. The situation is completely different for prolate spheroids,
which behave in the opposite way; for a population dominated by
prolate objects better results would be expected from application of
the standard method.

Finally, it would be important to assess the error involved in the
{\it measurement\/} of the ellipticity, but this would depend on a
number of conditions characterizing the specific set of observations
under investigation. Since the scale length of the lensed galaxies is
often smaller than $1 \arcsec$, seeing can be one important factor for
ground observations. Seeing makes galaxies rounder and thus leads to
an underestimate of the shear. Even if seeing effects can be partially
resolved by means of special algorithms, such as maximum-likelihood
and maximum-entropy image restoration (Lucy 1994; see also Bonnet \&
Mellier 1995, Kaiser {\it et al}. 1995, Luppino \& Kaiser 1997,
Villumsen 1995), the error on the measured ellipticity should be
larger for smaller galaxies. This is again in favor of the use of the
quadrupole method, which downplays the role of small galaxies. For
simplicity, seeing and other sources of error, such as Poisson noise,
sky luminosity, and pixeling are not considered in this paper and in
the simulations described below.

\titleb{Weight optimization}
The logarithmic choice of Eq.~(22) is one simple option compatible
with the requirements listed at the beginning of Sect.~3.2. One may
wonder whether it is possible to construct an optimal weight for a
given population of source galaxies. Suppose that the probability
distribution $p_Q\bigl( Q^{\rm s} \bigr)$ is known. Then, at least in
principle, one could try to find the weight function that minimizes
the error on $g$.

This interesting possibility is clarified by the following
example. Consider the weighted X method described by Eq.~(23) with a
weight function depending only on the determinant $D^{\rm s} = \det
Q^{\rm s}$. In this case the second requirement of Sect.~3.2, i.e.\
$W\bigl( k Q^{\rm s} \bigr) = f(k) W \big( Q^{\rm s} \bigr)$ just
demands that $W \bigl( D^{\rm s} \bigr)$ be a homogeneous
function. Therefore $W$ can be taken of the form $W\bigl( D^{\rm s}
\bigr) \propto \bigl( D^{\rm s} \bigr)^\eta$. Let us now introduce the
distribution probability $p_D \bigl( \chi^{\rm s}, D^{\rm s} \bigr) =
p_D \bigl( | \chi^{\rm s} |, D ^{\rm s} \bigr)$ for the source
galaxies. Obviously this distribution is related (in a complicated
way) to the quadrupole distribution $p_Q$. Using the relation $D^{\rm
s} = (\det J)^2 D$, where $D = \det Q$, and the form of $W\bigl(
D^{\rm s} \bigr)$, the variance of $g$ can be shown to be proportional
to the variance of $\widetilde{\chi^{\rm s}}$ divided by the square of
$W \bigl(\overline{D^{\rm s}} \bigr)$, where $\overline{D^{\rm s}}$ is
the mean value of $D^{\rm s}$ (see Eq.~(23) and the comment after
Eq.~(A7), $B$ being proportional to $W\bigl( \overline{D^{\rm s}}
\bigr)$). In other words the quantity to be minimized is
$$\eqalignno{
& {\Cov \bigl( \widetilde{\chi^{\rm s}} \bigr) \over \bigl[ W
\bigl(\overline{D^{\rm s}} \bigr) \bigr]^2} = { \pi \over N \bigl[ W \bigl(
\overline{D^{\rm s}} \bigr) \bigr]^2} \times {} & \cr
& \quad {} \times \int \bigl[ W \bigl(D^{\rm s} \bigr) \bigr]^2 |
\chi^{\rm s} |^3 p_D \bigl( |\chi^{\rm s}|, D^{\rm s} \bigr) \, \diff
| \chi^{\rm s} | \, \diff D^{\rm s} \; . & \autnum \cr}
$$
Recalling that $W\bigl( D^{\rm s} \bigr) \propto \bigl( D^{\rm s}
\bigr)^\eta$, we can find the value of $\eta$ that minimizes Eq.~(26)
provided that we know the source distribution $p_D$.

The value of $\eta$ that minimizes expression~(26) can be easily
obtained numerically. An approximate solution can be found by
expanding $D^{\rm s}$ near its mean value $\overline{D^{\rm s}}$. To
second order
$$
\eta_{\rm best} = {1 \over 4} - {\overline{D^{\rm s}} \bigl\langle
\bigl(D^{\rm s} - \overline{D^{\rm s}} \bigr) |\chi^{\rm s}|^2
\bigr\rangle \over 2 \bigl\langle \bigl( D^{\rm s} - \overline{D^{\rm
s}} \bigr)^2 |\chi^{\rm s}|^2 \bigr\rangle} \; . 
\eqno \autnum
$$
This should lead to the most accurate determination of $g$. The second
term of Eq.~(27), with its minus sign, takes into account the
correlation between the galaxy size $D^{\rm s}$, and its ellipticity
$\chi^{\rm s}$. If, as in our simulations, the largest galaxies are
the ones with smallest ellipticity, then this term is positive. In
fact, here it is convenient to use a large value for $\eta$ in order
to penalize smaller galaxies, which should be the ones with a large
ellipticity. Curiously, the case where size and ellipticity are fully
uncorrelated in the source population would still require a
non-trivial weight function ($\eta_{\rm best} = 1 / 4$).

Of course the weight optimization should be performed in {\it both\/}
the source size ($D^{\rm s}$) and ellipticity ($\chi^{\rm s}$). This
task is, obviously, much more difficult, mainly because of the
non-trivial dependence of $\chi$ on $\chi^{\rm s}$, given by
Eq.~(A8).

\titlea{Monte Carlo simulations}

\titleb{Source galaxies}
To test the various methods we have performed a series of Monte Carlo
simulations of measurements of the reduced shear parameter $g$, by
generating $N$ galaxies as a realization of two types of source
galaxies. At this stage it is not completely clear what is to be
considered a {\it realistic\/} population of source galaxies. Below we
focus on natural cases that we could draw from the literature.

The first galaxy distribution (Population~A) is characterized by a
source ellipticity distribution inferred from about $6000$ galaxies
observed in a single frame obtained at the $200$-inch Hale telescope
at Palomar (Brainerd {\it et al}.\ 1996). Calling $q$ the axis ratio
of the galaxy (with $0 \le q \le 1$), the empirical probability
distribution for $q$ is
$$
p_q (q) \simeq 64 q \exp ( - 8 q) \; .
\eqno \autnum
$$
The data by Oderwahn {\it et al}.\ (1997) basically confirm the
plausibility of this choice. The associated distribution for the
complex ellipticity $\chi^{\rm s}$ is shown in Fig.~1. The source
galaxies are largely dominated by disk galaxies, with significant
contributions from ellipticals. In order to specify the size
distribution, for simplicity we have followed Wilson {\it et al}.\
(1996) in adopting exponential luminosity profiles for all the source
objects, with scale-length $h$ uniformly distributed in the range
$[0.25, 0.65]$ arc-seconds. Finally, the position angle has been
chosen randomly with uniform distribution. We have assumed that the
three distributions (ellipticity, size, and position angle) are
independent. A correlation between ellipticity and size might be
present in a more realistic population of sources.  Based on this
choice of $p_q(q)$ we have $c
\simeq 0.0606$. Thus the expected error in the determination of $g$
for both the standard method and the simple quadrupole method is
$\approx 0.015/N$.

\begfig 6.9 cm
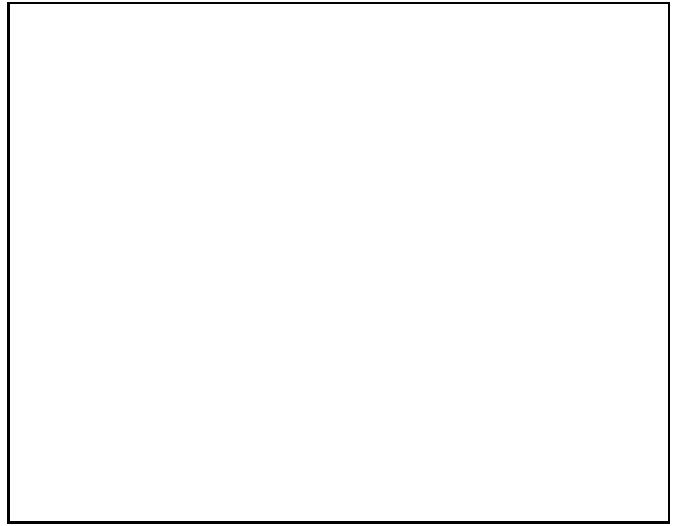
\figure{1}{The probability distribution of the source ellipticity for
the galaxy populations A and B described in the text. The distribution
is isotropic, i.e., $p\big(\chi^{\rm s} \big) = p\big( \big| \chi^{\rm
s} \big| \big)$. Note that in our notation the distribution is
normalized with the condition $2 \pi \int p \bigl( | \chi^{\rm s} |
\bigr) | \chi^{\rm s} | \, \diff | \chi^{\rm s} | = 1$.}
\endfig

The second simple population of source galaxies (Population~B) has
been generated by assuming that all galaxies are flat disks of the
same size and with random orientation (see also Bonnet \& Mellier
1995). A straightforward calculation shows that the probability
distribution for $q$ is uniform, i.e.
$$
p_q(q) = 1 \; .
\eqno \autnum
$$
In this case we have $c \simeq 0.2146$. The associated distribution
for $\chi^{\rm s}$ (see Fig.~1) shows a pronounced peak near $|\chi| =
1$. This distribution is not sharp in the sense of Sect.~2.3, because
of its relatively large value of $c$.

\titleb{Simulations and results}
The lens properties are then specified by means of the value of $g_0$;
the value of $(1 - \kappa)$ does not affect the results. For a given
lens, the observed quadrupole moment $Q^{(n)}$ of each source is
calculated by inverting Eq.~(8)
$$
Q^{(n)} = J^{-1} Q^{{\rm s}(n)} J^{-1} \; .
\eqno \autnum
$$
From the set of observed quadrupole moments $\bigl\{ \, Q^{(n)} \,
\bigr\}$ the methods X, Q, and W described in Sect.~2 and Sect.~3
allow us to determine the {\it measured\/} shear $g$. All methods
involve the resolution of an implicit equation (Eq.~(11), Eq.~(14),
and Eq.~(21)): this has been done with a simple Newton-Raphson
algorithm (see Press {\it et al}.\ 1992). For the results shown below
the W method has been implemented in the form of Eq.~(22); tests on
other types of weight functions, not shown here, have also been
performed.

\begfig 11 cm
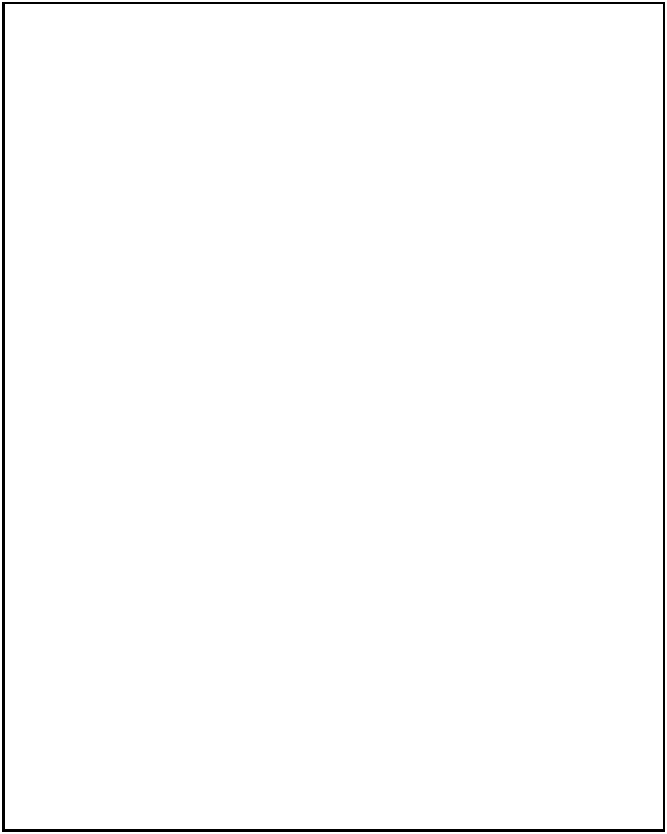
\figure{2}{The graphs show, for population~A, the mean error on $g$ as
a function of $g_0$. The error has been obtained from $10,000$
simulations ($N = 16$ galaxies). Frame (a) shows the results for the
standard method. The three methods are compared in frame (b). The
results agree very well with the $\bigl| 1 - |g_0|^2 \bigr|$ law.}
\endfig

The entire process has been repeated a large number of times
(typically $10,000$), each time based on a different realization of
the $N$ source galaxies. Thus for each method, the mean of the
measures of $g$ and the related errors (covariance matrix) have been
calculated.  The simulations immediately provide a simple check on the
mean of $\chi^{\rm s}$ and $Q^{\rm s}$ and on their covariance
matrices, consistent with the relation $c = e / M^2$ stated in
Sect.~3.1. Then we have compared the average errors of each method
with the expected error given by Eqs.~(13) and (19).

The main results for population~A of galaxies are shown in Fig.~2,
confirming the anticipated dependence on $|g_0|$; the expected
diagonal character of $\Cov_{ij}(g)$ for all the methods is also
recovered.

\begfig 9.14 cm
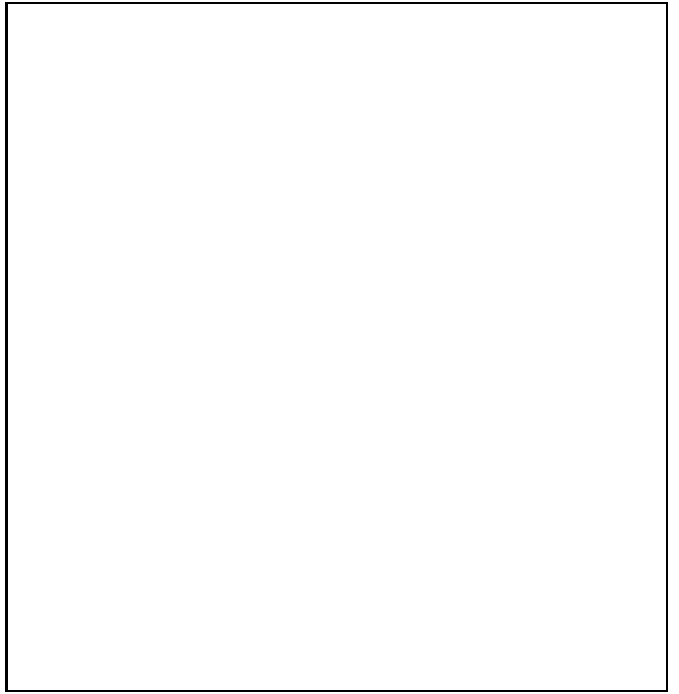
\figure{3}{The mean error on $g$ versus the number $N$ of galaxies
used in the reconstruction process. Symbols are the results of the
average error obtained with $10,000$ simulations for the three
specified methods, while the line is the $1/\sqrt{N}$ law given by
Eqs.~(13) and (19). The value of $g_0$ used is $g_0= 0.2 + 0.2
\imag$. The trends observed at $N \sim 30$ have been checked to
persist well beyond the range of $N$ shown here.} 
\endfig

As we can see from Fig.~3, the errors for the standard method (X)
follow the $1/\sqrt{N}$ law. They agree quite well with Eqs.~(13) and
(19) for all $g_0$ values. From these results on the standard method,
the galaxy distribution~A is found to be sufficiently sharp. The error
for the simple quadrupole method (Q) is instead about $15 \%$ lower
than in the standard method for every value of $g$ and $N$ (see the
short discussion given in Sect.~3.3). The weighted quadrupole method
(W) shows even smaller errors: these are about $28 \%$ smaller than
that of the standard method. The fact that the X symbols are closest
to the theoretical line just demonstrates that the asymptotic
approach at the basis of Eqs.~(13) and (19) is more ``natural'' for
the X method.

The results based on population~B are qualitatively similar. The
relevant ellipticity distribution is rather broad, and thus we expect
Eq.~(13) to give only approximate results. In fact, actual errors are
about $30 \%$ larger than those stated in Eq.~(13). For this galaxy
population the differences among the three methods are more
significant. The error in the simple quadrupole method is $27 \%$
smaller than that in the standard method, while in the weighted
quadrupole method it is $38 \%$ smaller.

In order to further test the sensitivity of our results to the
properties of the source galaxy distribution, we have run simulations
on another class of populations, not necessarily following empirical
suggestions (as was the case of Eq.~(28)), but rather mathematical
simplicity. This class of populations has been produced by imposing a
Gaussian distribution for $\chi^{\rm s}$ (suitably truncated at
$|\chi^{\rm s}| = 1$). One case, that we have called Population~C, is
characterized by a truncated Gaussian with covariance $c = 0.0606$ and
by the same size distribution as for population~A. Here we find that
the Q method performs better than the X method, with error smaller by
$\approx 20\%$; curiously, the W method implemented with Eq.~(22) here
leads to errors comparable to those of the X method. It thus appears
that the optimal choice of reconstruction method may be sensitive to
the detailed behavior of $p_q(q)$ around $q = 1$.

The specific examples addressed by the simulations can also be
interpreted in terms of an ``effective population'' of source
galaxies, determined by the adopted weights (which are a function of
$\chi^{\rm s}$). In other words, good weights act {\it as if\/} the
dispersion in the distribution of $\chi^{\rm s}$ were reduced. Of
course this is only an intuitive way of describing how the weights
operate in the process, since the resulting ``effective population''
depends on the unknown shear. In addition, one should also introduce
an ``effective $N$'' to describe the impact of the weights not
directly related to the shear.

\titlea{Conclusions}
In this paper we have analyzed in detail different methods aimed at
obtaining the reduced shear $g$ from the observed ellipticities of a
set of galaxies and we have focused on the issue of the accuracy of
such measurement. Monte Carlo simulations have demonstrated that
realistic distributions in general favor a new method introduced in
this paper, called the quadrupole method. In particular, the weighted
quadrupole method has been shown to perform best (the gain is of the
order of $\approx 15$--$30\%$ for the simple quadrupole method and can
be of the order of $\approx 30$--$40\%$ for the weighted quadrupole
method). This in turn leads to an improvement on the mass
determinations, which is anticipated to be by a similar factor.

\acknow{We thank Peter Schneider for many helpful discussions and
suggestions. This work has been partially supported by MURST and by
ASI of Italy.}

\appendix{A: The expected covariance matrix for a sharp ellipticity
distribution}

In order to calculate the covariance expected in a measurement of $g$
with the methods described in the main text, we use a general
expression known in the theory of errors (see Taylor 1982). Let $X$ be
a multidimensional random variable and let $Y = f(X)$. Suppose that
$f$ is a smooth function and that the probability distribution of $X$
is sharp, so that its covariance matrix $\Cov(X)$ is small. Then we
have (basically this is an application of the saddle-point integration
method)
$$\eqalignno{
\langle Y \rangle &\simeq f\bigl( \langle X \rangle \bigr) \; , &
\xautnum{A} \cr
\Cov(Y) &\simeq C \Cov(X) C^{\rm T} \; . & \xautnum{A} \cr}
$$
Here $C$ is the Jacobian matrix for $f$ calculated for the mean value of
$X$, i.e.
$$
C = \left. {\partial f \over \partial X} \right|_{X = \langle X
\rangle} \; .
\eqno \xautnum{A}
$$
If $Y$ is implicitly defined by $F(X, Y) = 0$, we can equally well
apply the previous relations using the inverse function theorem. In
this case the mean value of $Y$ is implicitly provided by $F\bigl(
\langle X \rangle, \langle Y \rangle \bigr) = 0$, while $C = - B^{-1} A$
with
$$\eqalignno{
A &= \left. {\partial F \over \partial X} \right|_{X = \langle X
\rangle, Y = \langle Y \rangle } \; , & \xautnum{A} \cr
B &= \left. {\partial F \over \partial Y} \right|_{X = \langle X
\rangle, Y = \langle Y \rangle} \; . & \xautnum{A} \cr}
$$
In conclusion in this case we have
$$\eqalignno{
& F\bigl( \langle X \rangle, \langle Y \rangle \bigr) \; \simeq 0 ,&
\xautnum{A} \cr
& \Cov(Y) \simeq \bigl( B^{-1} A \bigr) \Cov(X) \bigl( B^{-1} A
\bigr)^{\rm T} \; . & \xautnum{A} \cr}
$$
If we take $Y = \langle Y \rangle = \hbox{constant}$ and $X$ to be a
random variable, then we have $\Cov(F) = A \Cov(X) A^{\rm
T}$. Therefore, from Eq.~(A7) we may recognize that $\Cov(Y) \simeq
B^{-1} \Cov(F) \bigl(B^{-1}\bigr)^{\rm T}$.

Based on these expressions we can easily prove Eqs.~(12, 13) for the
standard method, and (18, 19) for the simple quadrupole method. For
example, for the standard method we will identify $Y$ with the reduced
shear and $F$ with $\tilde{\chi^{\rm s}}$, under the condition that
the lens be characterized by true reduced shear, $Y = g_0$.

\xtitleb{A}{Standard method}
In the standard method the reduced shear $g$ is calculated from
Eq.~(11). In principle this equation depends on $N$ complex random
variables $\chi^{(n)}$ and on the unknown reduced shear $g$. The
expected average $\langle \chi \rangle$ for the observed ellipticity
of every galaxy $\chi^{(n)}$ is easily obtained from the inverse of
Eq.~(9), that is
$$
\chi = {\chi^{\rm s} - 2 g + \chi^{\rm s *} g^2 \over 1 - 2 \Re
\bigl( \chi^{\rm s *} g \bigr) + |g|^2} \; .
\eqno \xautnum{A}
$$
By setting $\chi^{\rm s} = \bigl\langle \chi^{\rm s}
\bigr\rangle = 0$, this gives the mean
$$
\langle \chi \rangle = - {2 g_0 \over 1 + |g_0|^2} \; .
\eqno \xautnum{A}
$$
As in the main text, $g_0$ represents the {\it real\/} value of
$g$. We stress again that this equation is valid only in the case of a
sharp source ellipticity distribution ($c \ll 1$; see Schneider \&
Seitz 1995 for the relevant analysis).

The expected mean of measurements of $g$ obtained from Eq.~(11) obeys
the relation
$$
{1 \over N} \sum_{n = 1}^N \chi^{\rm s} \bigl( \langle \chi \rangle,
\langle g \rangle \bigr) = 0.
\eqno \xautnum{A}
$$
Using the calculated mean $\langle \chi \rangle$ of Eq.~(A9) above we
thus prove Eq.~(12).

Following Eq.~(A7), the covariance matrix for $g$ can be written
as
$$
\Cov(g) = \sum_{n=1}^N \bigl( B^{-1} A_n \bigr) \Cov(\chi) \bigl( B^{-1}
A_n \bigr)^{\rm T} \; .
\eqno \xautnum{A}
$$
The matrices $A_n$ and $B$ are the partial derivatives of $\tilde{\chi^{\rm
s}}(\chi, g)$:
$$\eqalignno{
A_n &= \left. {\partial \tilde{\chi^{\rm s}}\bigl( \bigl\{ \chi^{(m)}
\bigr\}, g \bigl) \over \partial \chi^{(n)}} \right|_{\chi^{(m)} =
\langle \chi \rangle, g = \langle g \rangle} & \cr
& = { 1 \over N} \left. {\partial \chi^{\rm s}\bigl( \chi^{(n)}, g
\bigr) \over \partial \chi^{(n)}} \right|_{\chi^{(n)} = \langle \chi
\rangle, g = \langle g \rangle} = {1 \over N} A \; , & \xautnum{A} \cr
B &= \left. {\partial \tilde{\chi^{\rm s}}\bigl( \bigl\{ \chi^{(m)}
\bigr\}, g \bigl) \over \partial g} \right|_{\chi = \langle \chi
\rangle, g = \langle g \rangle} & \cr
&= \left. {\partial \chi^{\rm s}(\chi, g) \over \partial g}
\right|_{\chi = \langle \chi \rangle, g = \langle g \rangle} \; . &
\xautnum{A} \cr}
$$
Note that the matrices $A_n = (1/N) A$ are all equal. The covariance
matrix $\Cov(\chi)$ can be expressed, by application of the inverse
function theorem, as
$$
\Cov(\chi) = A^{-1} \Cov\bigl( \chi^{\rm s} \bigr) \bigl( A^{-1}
\bigr)^{\rm T} = c A^{-1} \bigl( A^{-1} \bigr)^{\rm T} \; ,
\eqno \xautnum{A}
$$
so that Eq.~(A11) becomes
$$
\Cov(g) = {c \over N} B^{-1} \bigl( B^{-1} \bigl)^{\rm T} \; .
\eqno \xautnum{A}
$$
A simple algebraic calculation leads to
$$
B = {2 \over 1 - |g_0|^2} {\rm\ Id}
\eqno \xautnum{A}
$$
and thus we recover Eq.~(13).

\xtitleb{A}{Simple quadrupole method}
To show that the simple quadrupole method leads to similar results, we
start by considering the expected distribution for the observed
quadrupoles $Q$. From Eq.~(16) in the sharp distribution limit we have
$$
\langle Q \rangle = M J^{-2}
\eqno \xautnum{A}
$$
for every galaxy. Here $J^{-2}$ indicates $\bigl( J^{-1} \bigr)^2$,
i.e.\ the square of the inverse matrix of $J$. The expected average of
$g$ is calculated by solving Eq.~(14) with $g$ replaced by $\langle g
\rangle$, i.e.\
$$
\chi^{\rm s} \bigl( \langle Q \rangle, \langle g \rangle \bigr) = 0 \;
.
\eqno \xautnum{A}
$$
Since $\chi^{\rm s}(Q, g) \equiv \chi(JQJ)$, with $J$ any matrix with
associated reduced shear equal to $g$, we recover Eq.~(18).

In order to calculate the covariance matrix of $g$, we use
Eq.~(A7). From Eq.~(A2), the covariance of $\cal Q$ (recall the
definition $Q_{ij} = {\cal Q}_{i+j-1}$) can be written as
$$
\Cov({\cal Q}) = C \Cov \bigl( {\cal Q}^{\rm s} \bigr) C^{\rm T} \; ,
\eqno \xautnum{A}
$$
where $C$ is the Jacobian matrix
$$
C = \left. {\partial {\cal Q} \bigl( {\cal Q}^{\rm s}, g \bigr) \over
\partial {\cal Q}^{\rm s}} \right|_{{\cal Q}^{\rm s} = \langle {\cal
Q}^{\rm s} \rangle, g = \langle g \rangle} \; .
\eqno \xautnum{A}
$$
We observe that the argument of $\chi^{\rm s}$ used in Eq.~(14) is
$\tilde{\cal Q} = (1 / N) \sum_n {\cal Q}^{(n)}$: this allows us to
use $\tilde{\cal Q}$ as $X$ variable in the analysis. Thus the matrices
$A$ and $B$ of Eqs.~(A4) and (A5) are given by
$$\eqalignno{ A &= \left. {\partial \chi^{\rm s} \bigl( \tilde{\cal
Q}, g \bigr) \over \partial \tilde{\cal Q}} \right|_{\tilde{\cal Q} =
\langle {\cal Q} \rangle, g = \langle g \rangle} \; ,&
\xautnum{A} \cr 
B &= \left. {\partial \chi^{\rm s} \bigl( \tilde{\cal Q}, g \bigr) \over
\partial g} \right|_{\tilde{\cal Q} = \langle {\cal Q} \rangle, g =
\langle g \rangle} \; ,& \xautnum{A} \cr}
$$
where the average values of $\cal Q$ and $g$ are stated in Eqs.~(A17)
and (18). Note that the matrix $B$ defined here is the same as the one
used for the standard method and defined in Eq.~(A13). The covariance
matrix for $g$ is then
$$\eqalignno{
\Cov(g) &= \bigl( B^{-1} A \bigr) \Cov \biggl( {1 \over N} \sum_{n=1}^N
{\cal Q}^{(n)} \biggr) \bigl( B^{-1} A \bigr)^{\rm T} & \cr
&{} = {1 \over N} \bigl( B^{-1} A \bigr) C \Cov \bigl( {\cal Q}^{\rm s}
\bigr) C^T \bigl( B^{-1} A \bigr)^T \; . & \xautnum{A}}
$$
Therefore we have
$$
AC = {\partial \chi^{\rm s} \over \partial {\cal Q}} {\partial {\cal
Q} \over \partial {\cal Q}^{\rm s}} = \left. {\partial \chi^{\rm s}
\over \partial {\cal Q}^{\rm s}} \right|_{ {\cal Q}^{\rm s} = \langle
{\cal Q}^{\rm s} \rangle}
\eqno \xautnum{A}
$$
and thus
$$
AC \Cov\bigl( {\cal Q}^{\rm s} \bigr) (AC)^{\rm T} = \Cov\bigl(
\chi^{\rm s} \bigr) = c \hbox{ Id} \; .
\eqno \xautnum{A}
$$
Using Eq.~(A16) we finally obtain Eq.~(19).

\xtitleb{A}{The dependence on $g$}
From local observations it is not possible to choose between two
different solutions for the reduced shear related by $g'= 1 /
g^*$. We thus expect that the expression for the error will
incorporate this invariance property.

The $g \mapsto 1/g^*$ transformation is equivalent to a change of
sign in one eigenvalue of $J$ (see Schneider \& Seitz 1995), which is
not observable. Every reconstruction method is then bound to give two
solutions for the reduced shear.  As $g'$ is related to $g$, we can
calculate the expected error on $g'$ given the error on $g$:
$$
\Cov(g') = \left( {\partial g' \over \partial g} \right) \Cov(g)
\left( {\partial g' \over \partial g} \right)^{\rm T} \; .
$$
A simple calculation gives
$$
\left( {\partial g' \over \partial g} \right) = {1 \over |g|^4} \left(
\matrix{ g_1^2 - g_2^2 & -2 g_1 g_2 \cr -2 g_1 g_2 & g_2^2 - g_1^2
\cr } \right)
\eqno \xautnum{A}
$$
so that, as $\Cov(g) \propto \hbox{Id}$, we have
$$\eqalignno{
\Cov(g') = {1 \over |g|^4} \Cov(g) &= {c \bigl( 1/|g|^2 - 1 \bigr)^2
\over 4 N} \hbox{ Id} \cr
& {} = {c \bigl( 1 - |g'|^2 \bigr)^2 \over 4 N} \hbox{ Id} \; . &
\xautnum{A} \cr}
$$
Hence our expression for the error on $g$ reflects the $g \mapsto
1/g^*$ invariance.

\appendix{B: Basic structure of the simulation algorithm}
The simulation algorithm, briefly described in Sect.~4, is composed of
three different parts.

For a given galaxy distribution, we have generated $N$ source galaxies
(quadrupoles) using the transformation method (see Press {\it et al}.\
1992); the rejection method is less convenient here. Then we have
calculated, using Eq.~(30), the observed quadrupoles and
ellipticities.

Different methods (X, Q, and W) have been applied by solving
respectively Eqs.~(11), (14), and (21) with a simple Newton-Raphson
algorithm. This algorithm requires an initial point for the unknown
$g$. This has been chosen using Eq.~(A9). Through the entire
simulation algorithm we have supposed to be able to distinguish
between the two solutions $g$ and $1/g^*$. In practical cases this is
easy if the lens is non-critical. For critical lenses this ambiguity
can be resolved only globally. Note that this assumption has been
implicitly made throughout the paper (e.g., consider the justification
of Eqs.~(A10) and (A18)).

The previous steps have been repeated a large number of times with the
same $g_0$. For each individual ``simulation'' we have memorized the
calculated value of $g$. Finally, the mean and the covariance matrix
have been calculated from the data base of all the available results.

The algorithm has been implemented as a C++ code running on a IBM RISC
System/6000 590 machine.

\begref{References}
%
\ref Bartelmann M., Narayan R., Seitz S., Schneider P., 1996, ApJ 464, L115
%
\ref Bonnet H., Mellier Y., 1995, A\&A 303, 331
%
\ref Brainerd T.G., Blandford R.D., Smail I., 1996, ApJ 466, 623
\ref Gould A., Villumsen J., 1994, ApJ 428, 45
\ref Griffiths R.E., Casertano S., Im M., Ratnatunga K.U., 1996, MNRAS
282, 1159
%
\ref Kaiser N., Squires G., 1993, ApJ 404, 441
%
\ref Kaiser N., Squires G., Broadhurst T., 1995, ApJ 449, 460
%
\ref Kneib J.B., Soucail G., 1995, in: ``Astrophysical applications of
gravitational lensing'', Proc. IAU Symp.~173, ed.~C.S. Kochanek \&
J.N. Hewitt, Kluwer, Dordrecht, p.~129
%
\ref Kneib J.B., Ellis R.S., Smail I., Couch W.J., Sharples R.M., 1996, ApJ
471, 643.
%
\ref Lucy L.B., 1994, A\&A 289, 983
%
\ref Luppino G.A., Kaiser N., 1997, ApJ 475, 20
%
\ref Miralda-Escud\'e J., 1991, ApJ 370, 1
%
\ref Odewahn S.C., Burstein D., Windhorst R.A., 1997, ApJ, in press
%
\ref Press W.H., Teukolsky S.A., Vetterling W.T., Flannery B.P., 1992
Numerical Recipes in C, Cambridge University Press, Cambridge
%
\ref Schneider P., Ehlers J., Falco E.E., 1992, Gravitational Lenses,
Springer, Heidelberg
%
\ref Schneider P., Seitz C., 1995, A\&A 294, 411
%
\ref Schramm T., Kayser R., 1995, A\&A 299, 1
%
\ref Seitz C., Schneider P., 1995, A\&A 297, 287
\ref Taylor J.R., 1982, An Introduction to Error Analysis: The Study of
Uncertainties in Physical Measurements, University Science Books
%
\ref Tyson J.A., Valdes F., Jarvis J.F., Mills A.P., 1984, ApJ 281,
L59
%
\ref Webster R.L., 1985, MNRAS 213, 871
%
\ref Villumsen J.V., 1995, report MPA~880
%
\ref Wilson G., Cole S., Frenk C.S., 1996, MNRAS 280, 199
\endref

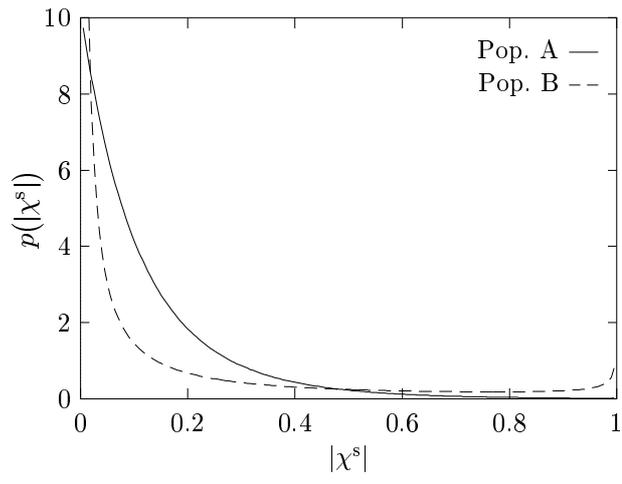
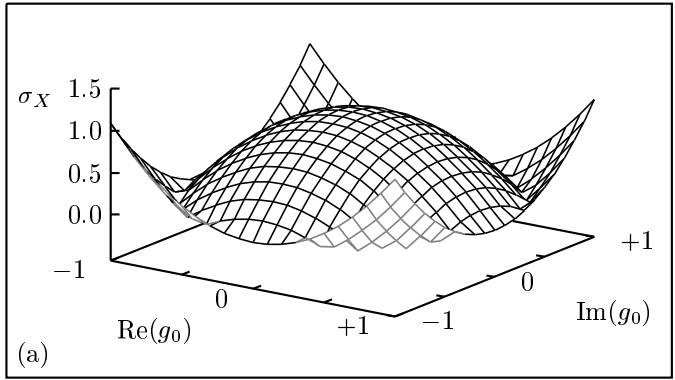
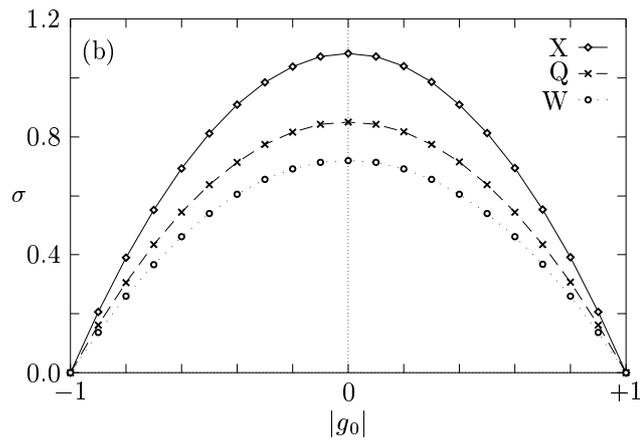
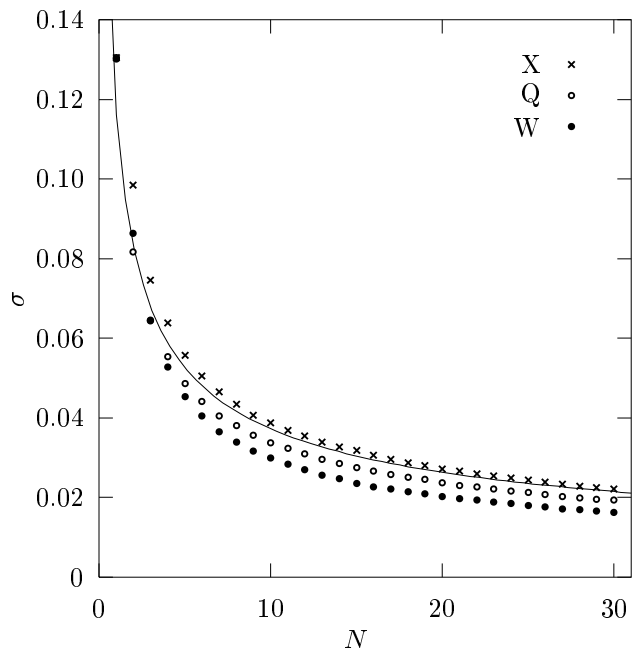
\bye